\documentclass[aps, reprint, superscriptaddress, twocolumn, prl, showpacs]{revtex4-1}

\pdfoutput=1

\usepackage{graphicx}
\usepackage{amsmath}
\usepackage{amssymb}
\usepackage{subfigure}
\usepackage{txfonts}
\usepackage{mathptmx}
\usepackage{hyperref}
\hypersetup{
    colorlinks,
    citecolor=red,
    filecolor=black,
    linkcolor=blue,
    urlcolor=magenta
}

\newcommand{\fref}[1]{Fig.~\ref{#1}}
\newcommand{\floor}[1]{{\lfloor\hspace{-.25mm} #1 \hspace{-.25mm}
    \rfloor}}

\allowdisplaybreaks[1]

\def\vp{p_\parallel}

\def\rhoo{\rho}
\def\rhoi{\rho_\odot}

\def\DMZ{D_\mathrm{MZ}}
\def\taur{\tau_\mathrm{R}}
\def\taub{\tau_\mathrm{F}}
\def\bfr{\mathbf{r}}
\def\bfn{\mathbf{n}}
\def\bfe{\mathbf{e}}
\def\bfv{\mathbf{p}}

\def\modvzero{p_0}

\begin{document}
\title{Machta-Zwanzig regime of anomalous diffusion in infinite-horizon billiards} 
\author{Giampaolo Cristadoro}
\affiliation{Dipartimento di Matematica, Universit\`a di Bologna, 
Piazza di Porta S. Donato 5, 40126 Bologna, Italy}
\author{Thomas Gilbert}
\affiliation{Center for Nonlinear Phenomena and Complex Systems,
  Universit\'e Libre  de Bruxelles, C.~P.~231, Campus Plaine, B-1050
  Brussels, Belgium}
\author{Marco Lenci}
\affiliation{Dipartimento di Matematica, Universit\`a di Bologna, 
Piazza di Porta S. Donato 5, 40126 Bologna, Italy}
\affiliation{Istituto Nazionale di Fisica Nucleare, Sezione di
  Bologna, Via Irnerio 46, 40126 Bologna, Italy}
\author{David P.~Sanders}
\affiliation{Departamento de F\'isica, Facultad de Ciencias, Universidad
Nacional Aut\'onoma de M\'exico,  Ciudad Universitaria,
04510 M\'exico D.F.,
 Mexico}

\begin{abstract}
We study diffusion on a periodic billiard table with infinite horizon
in the limit of narrow corridors. An effective trapping 
mechanism emerges according to which the process can be modeled
by a L\'evy walk combining exponentially-distributed trapping times
with free propagation along paths whose precise probabilities we
compute. This description yields an approximation of the mean squared
displacement of infinite-horizon billiards in terms of two transport
coefficients which generalizes to this anomalous regime the
Machta-Zwanzig approximation of normal diffusion in finite-horizon
billiards [Phys. Rev. Lett. \textbf{50}, 1959 (1983)]. 
\end{abstract}


\maketitle

The study of stochastic processes exhibiting anomalous transport has
attracted considerable attention in the recent past, in
particular in the context of continuous-time random walks
\cite{Haus:1987p513, Bouchaud:1990p1008, Weiss:1994RandomWalk,
  Shlesinger:1995uk, Benkadda:1998Chaos, Zaslavsky:2002Chaos}. It has
since evolved into  a highly interdisciplinary field, and includes a
large array of applications \cite{Klages:2008p11988}. 

From a fundamental perspective, a problem of specific interest
concerns the emergence of such stochastic processes from deterministic
dynamics \cite{Shlesinger:1993fw, Zaslavsky:1999Origin}, and the
derivation of their limiting distributions \cite{Klafter:1996PhysTod},
which typically differ from Gaussian, as, for instance, L\'evy stable
distributions \cite{gnedenko1968limit}.

Infinite-horizon periodic billiard tables, which are the focus
of this Letter, provide an example of mechanical systems whose
transport properties exhibit a weak form of anomalous behavior, in the
sense that there is a logarithmic correction to the linear growth in
time of the mean squared displacement \cite{Friedman1984p23,
  Bouchaud:1985p7680, Zacherl:1986p7768}; this is in fact a marginal
case, separating diffusive and superdiffusive regimes 
\cite{Geisel:1985p8023}. It is known, in particular, that the
distribution of the anomalously rescaled displacement vector,
$[\bfr(t) - \bfr(0)]/\sqrt{t \log t}$, converges weakly to
a normal distribution, whose covariance matrix is given in terms of a
scalar expression of the model's parameter, the 
so-called Bleher variance
\cite{Bleher:1992JSP315, Szasz:2007v129p59, Dolgopyat:2009p16456}.

As opposed to finite-horizon billiards, which are such that the
displacement between two successive collisions of a point 
particle with obstacles remains bounded, infinite-horizon billiards
admit, among their solutions, collisionless trajectories; 
point particles can therefore move arbitrarily far through infinite
\textit{corridors} devoid of obstacles. When the widths of these
corridors is taken to be vanishingly small, a separation between two
timescales occurs, the first associated with the propagation phases of
a trajectory across the billiard table along the corridors and the second with an
effective trapping mechanism which consists of a scattering phase that
takes place within a single billiard cell, separating successive
propagation phases. While the former timescale is independent of the
widths of the corridors, the latter grows unbounded in the
narrow-corridor regime. 

The purpose of this Letter is to show that, in the limit of narrow
corridors, a novel regime emerges, such that a
normal contribution to the finite-time diffusion coefficient may 
become arbitrarily larger than the anomalous component, even though
the latter diverges logarithmically as time increases while the former
remains asymptotically constant. 
Due to its connection with the Machta-Zwanzig approximation of the
diffusion coefficient of finite-horizon billiard tables
\cite{Machta:1983p182}, which  is obtained in a limit similar to that
of narrow corridors, we refer to this limit as a Machta-Zwanzig 
regime of anomalous diffusion in infinite-horizon billiards. 

To analyze this regime, we model the process on the infinite-horizon
billiard table by a L\'evy walk \cite{Klafter:1987Stochastic,
  Zumofen:1993p804}. Our approach is based on the framework of
continuous-time random walks and relies on the distinction between the
states of particles in propagating and scattering phases and the
derivation of a multistate generalized master equation
\cite{Kenkre:1973em,   Landman:1977PNAS430}. The process whereby a
walker on a lattice remains in a scattering phase during an
exponentially-distributed waiting time and then propagates in a random
direction over a free path of random length is such a multistate
process and it can be solved  analytically \cite{cristadoro:2014transport}.  
When applied to the infinite-horizon billiard, the probability
distribution of free paths decays algebraically with the third power
of their lengths \cite{Bouchaud:1985p7680, Bleher:1992JSP315}. The
mean squared displacement then has two relevant 
components: The first associated with the anomalous rescaling of the
displacement vector \cite{Bleher:1992JSP315, Szasz:2007v129p59,
  Dolgopyat:2009p16456}, and the second with a normal diffusion
component, similar to the Machta-Zwanzig approximation obtained in the
finite-horizon regime \cite{Machta:1983p182}. Whereas the former
contribution is second order in the widths of the corridors, the
latter is first order and, for finite time, becomes much
larger than the former.

We report numerical measurements of the mean square displacement of
infinite-horizon periodic billiard tables which exhibit those two
contributions, in agreement with the results predicted by the L\'evy
walk model. 

\begin{figure}[bht]
  \centering
  \includegraphics[width=.5\textwidth]{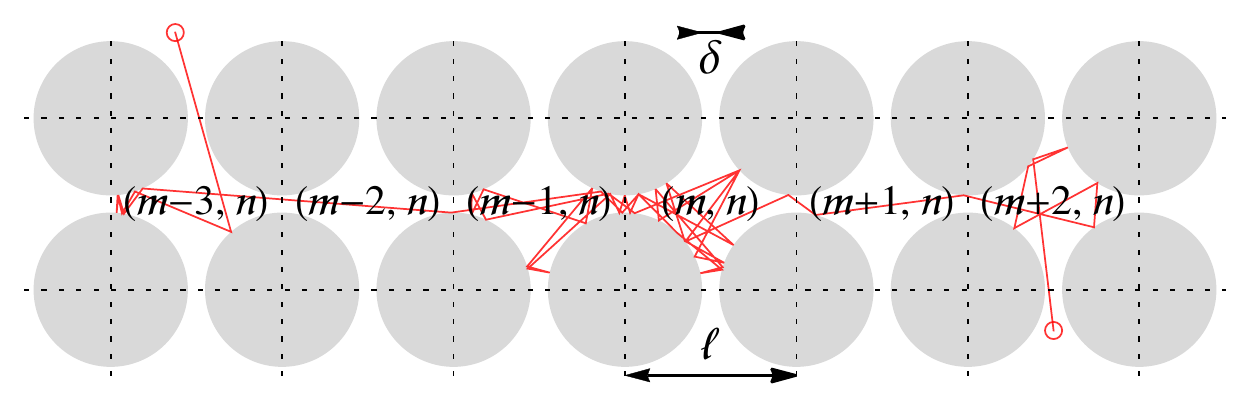}
  \caption{(Color online) Infinite-horizon periodic billiard
    table of square cells of sides $\ell$. The discs centered at the
    cells' corners have radii $\rhoo = 0.45\ell$. Infinite corridors
    of widths $\delta = \ell - 2\rhoo$ span along horizontal and 
    vertical axes, through the centers of each cell. The broken red
    line shows a trajectory.}
  \label{fig:table}
\end{figure}

\paragraph*{Model.}

We consider the simplest kind of billiard in the plane, which is
defined by a periodic array of square cells of sides 
$\ell$ with identical circular scatterers of radii $\rhoo$,
$0<\rhoo<\ell/2$, whose centers are placed at the corners of the
cells.  A point-particle moves freely on the exterior of these
obstacles, performing specular collisions upon their
boundaries; see \fref{fig:table}. 

This is an \emph{infinite-horizon} configuration, which refers to the
existence of collisionless trajectories along the vertical and
horizontal corridors that separate the obstacles. In 
contrast, a \emph{finite-horizon} configuration would occur if, for
instance, an additional circular obstacle of 
radius $\rhoi$, with $\ell/2 - \rhoo < \rhoi < \ell/\sqrt{2} - \rhoo$,
were placed at the center of each cell \footnote{Without loss of
  generality, we assume  $\rhoo\geq\rhoi$. For definiteness of a
  unique relevant timescale, Eq.~\eqref{eq:taur}, the separation
  between the central disc and the  outer ones should also be
  substantially larger than the separation between two outer discs.},
such that a particle would have to perform at least one collision in
each visited cell. The widths of the corridors is $\delta = 
\ell - 2\rhoo$, which we take to be a control parameter.

Transport of particles on such billiard tables can be studied in
terms of the time-evolution of the coarsegrained distribution of
particles in cell $\bfn\in\mathbb{Z}^2$, whose changes are determined
by the transitions particles make as they go from one cell to
another. Generally speaking, this is a complicated process which is
affected by  correlations between successive collision events. The
situation however simplifies when these correlations become
negligible, which occurs when $\delta \ll \ell$.  

\paragraph*{Infinite vs finite horizon.}

In finite-horizon tables, where diffusion is always normal
\cite{Bunimovich:1991p47}, this regime yields an approximation of the
process by a continuous-time random walk for displacements on the
lattice structure \cite{Gaspard:2012p26117}. The associated timescale
is given in terms of the residence time,  
\begin{equation}
  \taur = 
  \frac{\pi A}{4 \modvzero \delta},
  \label{eq:taur}
\end{equation}
where $A$ denotes the area of the elementary cell, \emph{viz}. $A =
\ell^2 - \pi (\rhoo^2 + \rhoi^2)$, and $\modvzero$ is the speed of
point particles \cite{Chernov:1997p1}.  

When $\taur$ is large with respect to the intercollisional time, a
particle typically spends a long time rattling about the same cell, 
making many collisions with its obstacles, before reaching one of the
four slits of widths $\delta$.  Under these conditions, the velocity
vectors of a particle entering and subsequently exiting a cell will
effectively be uncorrelated. The motion of a particle on the billiard
table can thus be approximated by a succession of scattering phases
with random waiting times, exponentially-distributed with scale
$\taur$, punctuated by instantaneous random hops from one cell to a
neighboring one.  

In this regime, the mean squared displacement grows linearly in time,
with coefficient approximated by $4\DMZ$, where 
\begin{equation}
  \DMZ = \frac{\ell^2}{4 \taur}
  \label{eq:difcoefFH}
\end{equation}
is a dimensional expression, known as the Machta-Zwanzig approximation
to the diffusion coefficient \cite{Machta:1983p182}. To leading order,
the actual diffusion coefficient differs from Eq. \eqref{eq:difcoefFH} by a
correction which is expected to be linear in $\delta/\ell$
\cite{Gilbert:2009p3207}. The validity of the Machta-Zwanzig
approximation thus relies on the 
separation $\delta\ll\ell$.  

In the case of infinite horizon, the presence of corridors renders the
Machta-Zwanzig regime more complex. In addition to the residence time
$\taur$,  Eq.~\eqref{eq:taur}, one has to take into consideration the 
timescale of propagation across a cell, $\taub$, which, in the
narrow-corridor regime is small with respect to $\taur$
\footnote{Strictly speaking, the residence time, Eq.~\eqref{eq:taur}, 
  accounts for the possibility of collisionless motion inside a
  cell. However in the narrow-corridor regime, the difference between
  $\taur$ and the actual residence time conditioned on particles
  performing collisions inside the cell is next order in the small
  parameter $\delta/\ell$ and will thus be neglected.},
\begin{equation}
  \taub \equiv \frac{\ell}{\modvzero} \ll \taur.
  \label{eq:septimes}
\end{equation} 
When the phase-space coordinates of a particle are such that it
crosses the boundary between two cells with velocity almost
perpendicular to it, the next scattering phase may take place at a
remote distance, after a time approximately equal to $\taub$
multiplied by the number of cells traveled free of collisions.

The assumption that correlations between successive scattering phases
become negligible amounts to approximating the transport process on
the infinite-horizon billiard table by a L\'evy walk, whose
distribution of free paths, i.e.~the distance a particle propagates free
of collisions, matches that of the infinite-horizon
billiard, taking into account the  timedelay induced by the
propagation along free paths. The general framework for analyzing
L\'evy walks where a distinction occurs between the states of
particles in scattering and propagating phases was described in
Ref.~\cite{cristadoro:2014transport}. Whereas particles in a
scattering state make transitions at random times, exponentially
distributed with scale $\taur$, and move in a random direction to a
neighboring cell, the transitions of particles in a propagating state
are deterministic: They take place after exactly time $\taub$ and are
accompanied by displacements along the same direction as the previous
transition. 

\paragraph*{Parameter values.} 

We label the internal state of a walker by $k \in \mathbb{Z}$, with $k=0$
denoting the scattering state and $k\geq 1$ a propagating state, where
$k$ corresponds to the total number of steps of duration $\taub$ until the
propagating state returns to a scattering one. The direction of
propagation is fixed in the propagating state. 
When the particle is in the scattering state, we let $\mu_k$ denote the
probability of a transition to either a scattering state, 
$k=0$, or a propagating state, $k\geq 1$, for any of the
four lattice directions the walker can move in. 

In the narrow-corridor regime, the overwhelming majority of
transitions are from scattering to scattering states; only rarely do
particles perform  transitions from scattering to propagating
states. When they do, however, as mentioned earlier, the probability
of a long excursion to a propagating state, $k\gg1$, decays with the
third power of their lengths, $\mu_k \sim k^{-3}$. In fact, to first
order in $\delta/\ell$, we can write  
\begin{equation}
  \mu_k = 
  \begin{cases}
    1 - \frac{\delta}{4\ell},& k = 0,\\
    \frac{\delta}{k(k+1)(k+2)\ell},& k \geq 1.
  \end{cases}
  \label{eq:rhok}
\end{equation}

\begin{figure}[t]
  \centering
  \includegraphics[width=.4\textwidth]{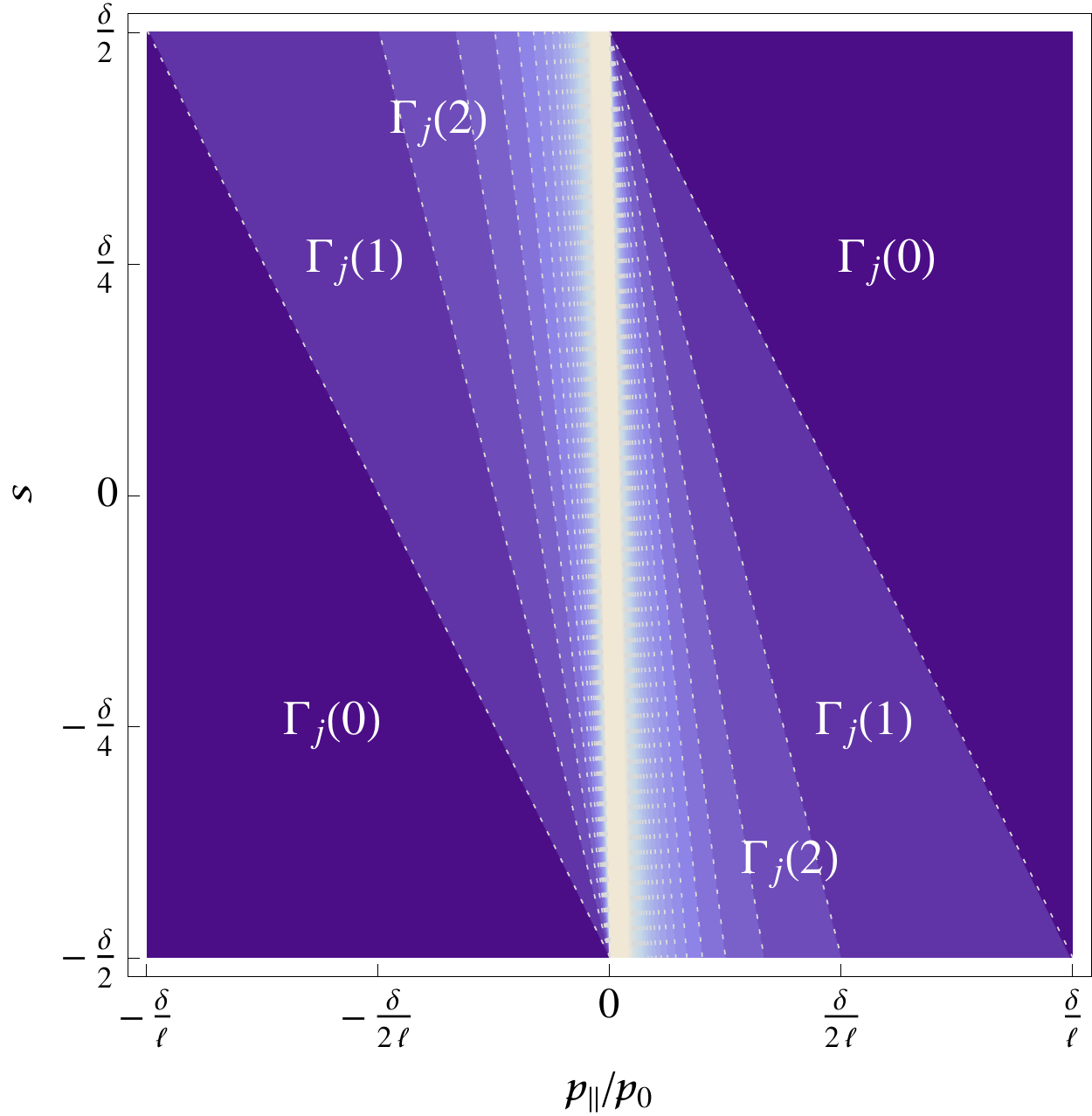}
  \caption{(Color online) Graph showing two-dimensional phase-space
    sets according to the distance separating phase points at the
    boundary between two cells from their next collision on a
    disc. The darkest color encodes sets of points whose next
    collision takes place in the cell they move into; the brighter the
    colors the more distant the next collision. The two axes
    correspond, respectively, to $\vp/\modvzero$, which denotes the
    fraction of the velocity component directed along the boundary,
    $-1 < \vp/\modvzero < 1$ (here restricted to $-\delta/\ell <
    \vp/\modvzero < \delta/\ell$) and to  the position $s$ along the
    boundary,  $-\delta/2 < s < \delta/2$, where $\delta = \ell -
    2\rhoo$ ($\rhoo = 0.45\ell$), is the width of the  corridor. With
    these coordinates, the area corresponds to the natural invariant
    measure, which is normalized if divided by $2\delta$.}
  \label{fig:histo}
\end{figure}

To obtain this result, consider the phase-space sets that lie at the
boundary between two neighboring cells $\bfn$ and $\bfn 
+ \bfe_j$, where $\bfe_j$ is a unit vector in one of the
four lattice directions, $j\in\{1,\dots,4\}$. A particle in cell $\bfn$
which crosses over to cell $\bfn + \bfe_j$ is mapped at this
boundary to a phase point with coordinate $|s| <  \delta/2$, along
the direction of the separation between the two cells, with velocity
$\bfv$ such that $\bfv\cdot\bfe_j \equiv p_\perp > 0$. 

Let $\Gamma_j(k)$ denote the set of such phase points which
are mapped by the flow to the next collision event on an obstacle in
cell $\bfn +k\,\bfe_j$. Since the points of these sets can be mapped
back to their preceding collision on an obstacle in cell $\bfn - l\,
\bfe_j$, for some  $l\geq0$, they are associated with point particles
in a propagation phase of length at least $k$. 
We must measure these points by means of the natural invariant
measure, the one induced by the Liouville measure on the
cross-sections defined by collisions or cell crossings, which is
known to be the area in phase space as parameterized in
\fref{fig:histo}. Therefore, up to a normalization factor, the area of
$\Gamma_j(k)$ is given by $\sum_{i = k}^\infty  \mu_i$. 
By geometric arguments, and as can also be seen 
from \fref{fig:histo}, the measure of
$\cup_{i=k}^\infty \Gamma_j(i)$ is, to leading order in $\delta/\ell$,
proportional to the area of  a right triangle of base 
$\delta/(k\ell)$. We therefore have, for $k\geq 1$, 
\begin{equation}
\sum_{i = k}^\infty \sum_{j   =
  i}^\infty \mu_j = \frac{\delta}{2 k \ell}, 
\end{equation}
which implies, and thus justifies, Eq.~\eqref{eq:rhok}.

\paragraph*{Anomalous diffusion.} 

We let $\bfr = \bfn\ell$ denote the displacement of L\'evy walkers on
the two-dimensional square lattice, measured from the origin where
they are initially located, and obtain an expression of their
mean squared displacement as a function of time, $\langle r^2
\rangle_t$, in terms of the time-integral of the overall fraction of
particles in a scattering state, $\sigma(t)$,  
\begin{equation}
  \frac{\langle r^2 \rangle_t}{\ell^2}
  = \frac{1}{\taur}
  \int_0^{t} \mathrm{d} s\, 
  \sigma(s) + 
  \frac{\delta}{2 \ell\taur}
  \sum_{k=1}^\floor{t/\taub}
  \frac{2 k + 1}{k(k+1)}
  \int_0^{t - k\taub} \mathrm{d} s\, \sigma(s);
 \label{eq:msdexact}
\end{equation}
see Ref.~\cite{cristadoro:2014transport} for further details.
In the stationary state,  the fraction of particles in a scattering
state is simply given by the ratio of the average time spent in the
scattering state, $\taur$, to the average return time to this state,
$\taur + \sum_{k=1}^\infty k \mu_k \taub$. Substituting the transition
probabilities, Eq.~\eqref{eq:rhok}, yields 
\begin{equation}
  \lim_{t\to\infty} \sigma(t) = 
  \frac{\taur}{\taur + \frac{\delta}{2\ell} \taub}  
  \simeq 1 - \frac{\delta}{2\ell} \frac{\taub}{\taur}.
\end{equation}
Plugging this expression into Eq.~\eqref{eq:msdexact}, we obtain an
expression of the mean square displacement of walkers in terms of
harmonic numbers, which, in the long-time limit, reduces to 
\begin{equation}
  \frac{\langle r^2 \rangle_t }{4t} 
  = \frac{\ell^2}{4\taur} + \frac{\delta\ell}{4\taur} \left[
  \log t + \mathcal{O}(1)
  \right].
  \label{eq:msdlargetime}
\end{equation}
This is our main result. It generalizes to infinite-horizon billiard
tables in the narrow-corridor regime the Machta-Zwanzig
approximation of the diffusion coefficient for finite-horizon tables, 
Eq.~\eqref{eq:difcoefFH}.

For short times, the right-hand side of Eq.~\eqref{eq:msdlargetime} is
dominated by the constant term, which, to leading order in $\delta/\ell$,
corresponds to a regime of normal diffusion, with the coefficient
\eqref{eq:difcoefFH}, consistent with the Machta-Zwanzig approximation
of the (normal) diffusion coefficient. 

The term carrying the logarithmic correction has coefficient
$\delta \ell/(4\taur)$, which is identical to the Bleher limiting
variance of the anomalously rescaled limiting distribution
\cite{Bleher:1992JSP315, Szasz:2007v129p59,
  Dolgopyat:2009p16456}, i.e.~such that the displacement vector
rescaled by the square root of ${t\log t}$ converges in distribution
to a centered normal distribution whose covariance matrix reduces to a
scalar given by this coefficient \footnote{The second moment actually
  scales with coefficient given by twice the Gaussian
  value. See discussion in Ref.~\cite{cristadoro:2014measuring}. }. 

Whereas the coefficient of the constant term on the right-hand side of
Eq.~\eqref{eq:msdlargetime} is first order in the small parameter
$\delta/\ell$, the coefficient of the logarithmic term is second
order. Such a contribution thus becomes significant only for times
such that $\log t \approx \ell/\delta$. In the narrow-corridor regime,
however, the constraints on the integration times are such that $\log
t$ remains small with respect $\ell/\delta$. 

\begin{figure}[htb]
  \centering
  \includegraphics[width=.45\textwidth]
  {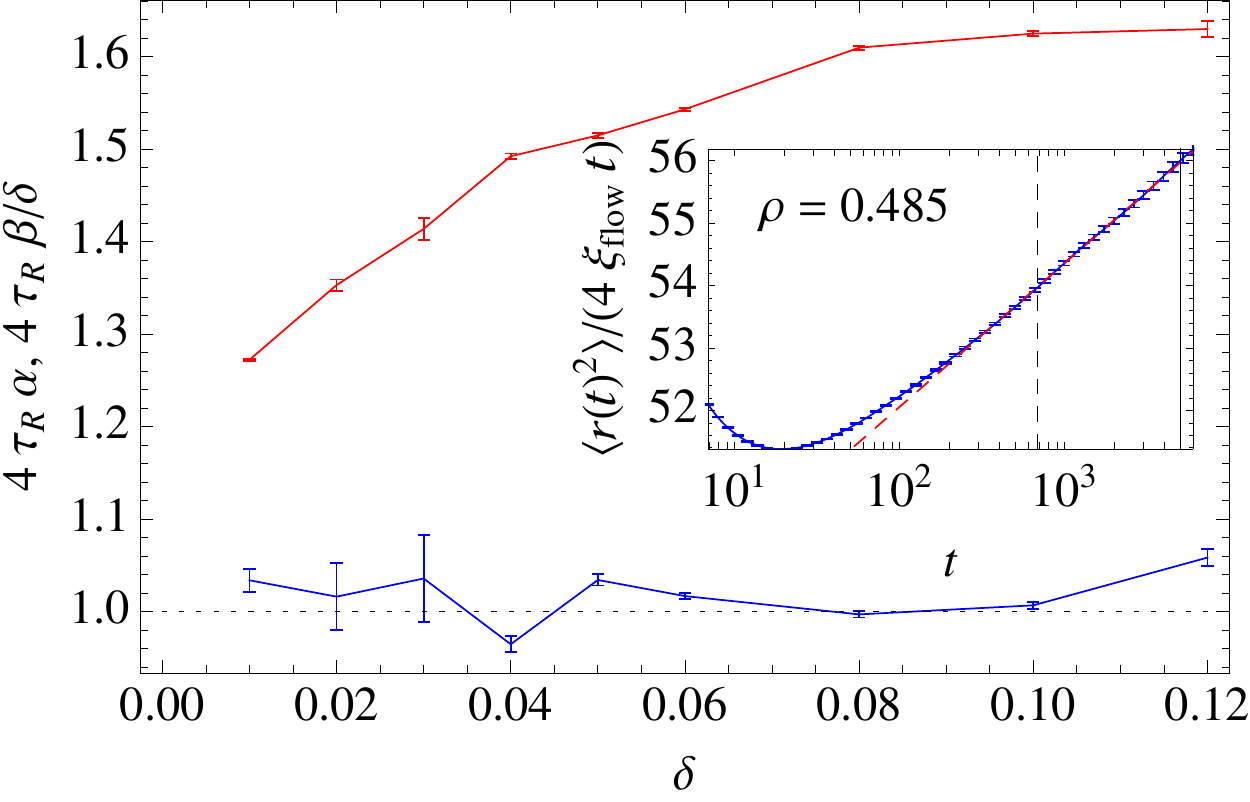}
  \caption{(Color online) Computed slopes, $\beta$ (blue, lower
    curve), and intercepts, $\alpha$ (red, upper curve),
    of the mean squared displacement of point particles on the
    infinite-horizon Lorentz gas, fitted according to Eq.~\eqref{eq:msdfit}
    for different values of the parameter $\delta$. The values found
    are here normalized by those predicted by
    Eqs.~\eqref{eq:alpha} and \eqref{eq:beta}. To illustrate the
    fitting procedure,  the inset shows the mean squared displacement
    (blue curve, including error bars) measured, as a function of
    time, for $\delta = 0.03$. The red dashed line is the result of a
    linear fit performed in the interval marked by the two vertical
    lines; see Ref.~\cite{cristadoro:2014measuring} for further
    details on this procedure. The units are chosen such that $\ell
    \equiv \modvzero \equiv 1$.
  }
  \label{fig:msqdisp}
\end{figure}

\paragraph*{Numerical results.}

Following the results presented in
Ref.~\cite{cristadoro:2014measuring}, we perform numerical 
measurements of the mean squared displacement of infinite-horizon
billiard tables such as shown in \fref{fig:table} 
and determine a range of time values such that the distribution of
free paths is accurately sampled, which, according to 
Eq.~\eqref{eq:rhok}, scales like the square root of the total number of
initial conditions taken (typically $10^9$). In that range, we
seek an asymptotically affine fitting function of $\log t$ for the normally
rescaled mean squared displacement, 
\begin{equation}
  \frac{\langle r^2 \rangle_t }{4t} 
  \sim \alpha + \beta \log t,
  \label{eq:msdfit}
\end{equation}
where $\alpha$ and $\beta$ are implicit functions of time, and are
expected to converge to the values predicted by
Eq.~\eqref{eq:msdlargetime} as $t\to\infty$,
i.e.
\begin{align}
  \lim_{\delta\to0} \lim_{t\to\infty} \frac{4 \taur \alpha}{\ell^2} = 1,
  \label{eq:alpha}\\
  \lim_{t\to\infty} \frac{4 \taur \beta}{\delta \ell} = 1,
  \label{eq:beta}
\end{align}
where, in the first line, the narrow-corridor limit takes care of
$\delta$-dependent corrections to $\alpha$ our theory does not
account for.   

Values found for these fitting parameters are reported in
\fref{fig:msqdisp} for different values of $\delta$. For the parameter
$\beta$, on the one hand, one expects Eq.~\eqref{eq:beta} to hold
 for all values of $\delta$ in the range shown and, in view of the
difficulties presented by such measurements
\cite{cristadoro:2014measuring}, the agreement is indeed rather good,
especially given the prevalence of finite-time effects when
$\delta\to0$. There is, on the other hand, no analytic prediction for
the value of the parameter $\alpha$, other than that given in the
narrow-corridor limit, Eq.~\eqref{eq:alpha}. Nevertheless, our data
provides clear evidence in support of this result.  

We should note that, in contrast to the approximating L\'evy walk, for
which corrections to the zeroth order result \eqref{eq:msdlargetime} are
found to be negative, the corrections to the zeroth order result for
measurements performed for billiards appear to be positive at first
order in $\delta/\ell$. This is to be expected, since memory effects
should indeed bring about corrections of the same order, as is the
case with finite-horizon billiards \cite{Gilbert:2009p3207}; such
corrections may well predominate.

\paragraph*{Conclusions}

Infinite-horizon billiard tables with narrow corridors display
anomalous transport properties such that the logarithmic divergence in
time of the mean squared displacement must effectively be treated as a
subleading contribution with respect to a normally diffusive one. 

Our stochastic analysis of the process in terms of a L\'evy walk with
both scattering phases, characterized by random waiting times with
exponential distributions, and propagating phases along the table's
corridors provides two quantitative predictions 
which match the Machta-Zwanzig dimensional prediction of the (normal)
diffusion coefficient, on the one hand, and the Bleher variance of the
anomalously rescaled process, on the other hand. Their physical
interpretations is, moreover, transparent: (i) the
overwhelming majority of transitions taking place on the billiard
table are similar to those observed in finite-horizon billiard tables,
giving rise to the predominant normal contribution to the mean 
squared displacement, and (ii) rare scattering events allow
propagation along the billiard's corridors over long distrances whose
lengths follow a precise distribution, at the origin of the anomalous
contribution to the  mean squared displacement. As our numerical
results make clear, ignoring the first of these two contributions
would obstruct the accurate measurement of the second.

We conclude by observing that the scaling properties of the transition
probabilities, Eq.~\eqref{eq:rhok}, can be generalized to other
values, extending the relevance of our approach well beyond the regime
studied in this Letter. As discussed in
Ref.~\cite{cristadoro:2014transport}, tuning the parameter values 
allows the description of both normal and anomalous transport regimes,
including ballistic transport.
Further applications will be reported elsewhere.

\begin{acknowledgments}
This work was partially supported by FIRB-Project No. RBFR08UH60
(MIUR, Italy), by SEP-CONACYT Grant No. CB-101246 and DGAPA-UNAM
PAPIIT Grant No. IN117214 (Mexico), and by FRFC convention 2,4592.11
(Belgium). T.G. is financially supported by the (Belgian) FRS-FNRS.   
\end{acknowledgments}


\begin{thebibliography}{32}%
\makeatletter
\providecommand \@ifxundefined [1]{%
 \@ifx{#1\undefined}
}%
\providecommand \@ifnum [1]{%
 \ifnum #1\expandafter \@firstoftwo
 \else \expandafter \@secondoftwo
 \fi
}%
\providecommand \@ifx [1]{%
 \ifx #1\expandafter \@firstoftwo
 \else \expandafter \@secondoftwo
 \fi
}%
\providecommand \natexlab [1]{#1}%
\providecommand \enquote  [1]{``#1''}%
\providecommand \bibnamefont  [1]{#1}%
\providecommand \bibfnamefont [1]{#1}%
\providecommand \citenamefont [1]{#1}%
\providecommand \href@noop [0]{\@secondoftwo}%
\providecommand \href [0]{\begingroup \@sanitize@url \@href}%
\providecommand \@href[1]{\@@startlink{#1}\@@href}%
\providecommand \@@href[1]{\endgroup#1\@@endlink}%
\providecommand \@sanitize@url [0]{\catcode `\\12\catcode `\$12\catcode
  `\&12\catcode `\#12\catcode `\^12\catcode `\_12\catcode `\%12\relax}%
\providecommand \@@startlink[1]{}%
\providecommand \@@endlink[0]{}%
\providecommand \url  [0]{\begingroup\@sanitize@url \@url }%
\providecommand \@url [1]{\endgroup\@href {#1}{\urlprefix }}%
\providecommand \urlprefix  [0]{URL }%
\providecommand \Eprint [0]{\href }%
\providecommand \doibase [0]{http://dx.doi.org/}%
\providecommand \selectlanguage [0]{\@gobble}%
\providecommand \bibinfo  [0]{\@secondoftwo}%
\providecommand \bibfield  [0]{\@secondoftwo}%
\providecommand \translation [1]{[#1]}%
\providecommand \BibitemOpen [0]{}%
\providecommand \bibitemStop [0]{}%
\providecommand \bibitemNoStop [0]{.\EOS\space}%
\providecommand \EOS [0]{\spacefactor3000\relax}%
\providecommand \BibitemShut  [1]{\csname bibitem#1\endcsname}%
\let\auto@bib@innerbib\@empty
\bibitem [{\citenamefont {Haus}\ and\ \citenamefont
  {Kehr}(1987)}]{Haus:1987p513}%
  \BibitemOpen
  \bibfield  {author} {\bibinfo {author} {\bibfnamefont {J.~W.}\ \bibnamefont
  {Haus}}\ and\ \bibinfo {author} {\bibfnamefont {K.~W.}\ \bibnamefont
  {Kehr}},\ }\href@noop {} {\bibfield  {journal} {\bibinfo  {journal} {Physics
  Reports}\ }\textbf {\bibinfo {volume} {150}},\ \bibinfo {pages} {263}
  (\bibinfo {year} {1987})}\BibitemShut {NoStop}%
\bibitem [{\citenamefont {Bouchaud}\ and\ \citenamefont
  {Georges}(1990)}]{Bouchaud:1990p1008}%
  \BibitemOpen
  \bibfield  {author} {\bibinfo {author} {\bibfnamefont {J.-P.}\ \bibnamefont
  {Bouchaud}}\ and\ \bibinfo {author} {\bibfnamefont {A.}~\bibnamefont
  {Georges}},\ }\href@noop {} {\bibfield  {journal} {\bibinfo  {journal}
  {Physics Reports}\ }\textbf {\bibinfo {volume} {195}},\ \bibinfo {pages}
  {127} (\bibinfo {year} {1990})}\BibitemShut {NoStop}%
\bibitem [{\citenamefont {Weiss}(1994)}]{Weiss:1994RandomWalk}%
  \BibitemOpen
  \bibfield  {author} {\bibinfo {author} {\bibfnamefont {G.~H.}\ \bibnamefont
  {Weiss}},\ }\href@noop {} {\emph {\bibinfo {title} {{Aspects and Applications
  of the Random Walk}}}}\ (\bibinfo  {publisher} {North-Holland},\ \bibinfo
  {address} {Amsterdam},\ \bibinfo {year} {1994})\BibitemShut {NoStop}%
\bibitem [{\citenamefont {Shlesinger}\ \emph {et~al.}(1995)\citenamefont
  {Shlesinger}, \citenamefont {Zaslavsky},\ and\ \citenamefont
  {Frisch}}]{Shlesinger:1995uk}%
  \BibitemOpen
  \bibinfo {editor} {\bibfnamefont {M.~F.}\ \bibnamefont {Shlesinger}},
  \bibinfo {editor} {\bibfnamefont {G.~M.}\ \bibnamefont {Zaslavsky}}, \ and\
  \bibinfo {editor} {\bibfnamefont {U.}~\bibnamefont {Frisch}},\ eds.,\
  \href@noop {} {\emph {\bibinfo {title} {{L{\'e}vy Flights and Related Topics
  in Physics}}}},\ \bibinfo {series} {Lecture Notes in Physics}, Vol.\ \bibinfo
  {volume} {450}\ (\bibinfo  {publisher} {Springer},\ \bibinfo {address}
  {Berlin, Heidelberg},\ \bibinfo {year} {1995})\BibitemShut {NoStop}%
\bibitem [{\citenamefont {Benkadda}\ and\ \citenamefont
  {Zaslavsky}(1998)}]{Benkadda:1998Chaos}%
  \BibitemOpen
  \bibinfo {editor} {\bibfnamefont {S.}~\bibnamefont {Benkadda}}\ and\ \bibinfo
  {editor} {\bibfnamefont {G.~M.}\ \bibnamefont {Zaslavsky}},\ eds.,\
  \href@noop {} {\emph {\bibinfo {title} {{Chaos, Kinetics and Nonlinear
  Dynamics in Fluids and Plasmas}}}},\ \bibinfo {series} {Lecture Notes in
  Physics}, Vol.\ \bibinfo {volume} {511}\ (\bibinfo  {publisher} {Springer},\
  \bibinfo {address} {Berlin Heidelberg},\ \bibinfo {year} {1998})\BibitemShut
  {NoStop}%
\bibitem [{\citenamefont {Zaslavsky}(2002)}]{Zaslavsky:2002Chaos}%
  \BibitemOpen
  \bibfield  {author} {\bibinfo {author} {\bibfnamefont {G.~M.}\ \bibnamefont
  {Zaslavsky}},\ }\href@noop {} {\bibfield  {journal} {\bibinfo  {journal}
  {Physics Reports}\ }\textbf {\bibinfo {volume} {371}},\ \bibinfo {pages}
  {461} (\bibinfo {year} {2002})}\BibitemShut {NoStop}%
\bibitem [{\citenamefont {Klages}\ \emph {et~al.}(2008)\citenamefont {Klages},
  \citenamefont {Radons},\ and\ \citenamefont {Sokolov}}]{Klages:2008p11988}%
  \BibitemOpen
  \bibfield  {author} {\bibinfo {author} {\bibfnamefont {R.}~\bibnamefont
  {Klages}}, \bibinfo {author} {\bibfnamefont {G.}~\bibnamefont {Radons}}, \
  and\ \bibinfo {author} {\bibfnamefont {I.~M.}\ \bibnamefont {Sokolov}},\
  }\href@noop {} {\emph {\bibinfo {title} {{Anomalous transport: Foundations
  and applications}}}}\ (\bibinfo  {publisher} {Wiley-VCH Verlag},\ \bibinfo
  {address} {Weinheim},\ \bibinfo {year} {2008})\BibitemShut {NoStop}%
\bibitem [{\citenamefont {Shlesinger}\ \emph {et~al.}(1993)\citenamefont
  {Shlesinger}, \citenamefont {Zaslavsky},\ and\ \citenamefont
  {Klafter}}]{Shlesinger:1993fw}%
  \BibitemOpen
  \bibfield  {author} {\bibinfo {author} {\bibfnamefont {M.~F.}\ \bibnamefont
  {Shlesinger}}, \bibinfo {author} {\bibfnamefont {G.~M.}\ \bibnamefont
  {Zaslavsky}}, \ and\ \bibinfo {author} {\bibfnamefont {J.}~\bibnamefont
  {Klafter}},\ }\href@noop {} {\bibfield  {journal} {\bibinfo  {journal}
  {Nature}\ }\textbf {\bibinfo {volume} {363}},\ \bibinfo {pages} {31}
  (\bibinfo {year} {1993})}\BibitemShut {NoStop}%
\bibitem [{\citenamefont {Zaslavsky}(1999)}]{Zaslavsky:1999Origin}%
  \BibitemOpen
  \bibfield  {author} {\bibinfo {author} {\bibfnamefont {G.~M.}\ \bibnamefont
  {Zaslavsky}},\ }\href@noop {} {\bibfield  {journal} {\bibinfo  {journal}
  {Physics Today}\ }\textbf {\bibinfo {volume} {52}},\ \bibinfo {pages} {39}
  (\bibinfo {year} {1999})}\BibitemShut {NoStop}%
\bibitem [{\citenamefont {Klafter}\ \emph {et~al.}(1996)\citenamefont
  {Klafter}, \citenamefont {Shlesinger},\ and\ \citenamefont
  {Zumofen}}]{Klafter:1996PhysTod}%
  \BibitemOpen
  \bibfield  {author} {\bibinfo {author} {\bibfnamefont {J.}~\bibnamefont
  {Klafter}}, \bibinfo {author} {\bibfnamefont {M.~F.}\ \bibnamefont
  {Shlesinger}}, \ and\ \bibinfo {author} {\bibfnamefont {G.}~\bibnamefont
  {Zumofen}},\ }\href@noop {} {\bibfield  {journal} {\bibinfo  {journal}
  {Physics Today}\ }\textbf {\bibinfo {volume} {49}},\ \bibinfo {pages} {33}
  (\bibinfo {year} {1996})}\BibitemShut {NoStop}%
\bibitem [{\citenamefont {Gnedenko}\ and\ \citenamefont
  {Kolmogorov}(1968)}]{gnedenko1968limit}%
  \BibitemOpen
  \bibfield  {author} {\bibinfo {author} {\bibfnamefont {B.~V.}\ \bibnamefont
  {Gnedenko}}\ and\ \bibinfo {author} {\bibfnamefont {A.~N.}\ \bibnamefont
  {Kolmogorov}},\ }\href@noop {} {\emph {\bibinfo {title} {{Limit Distributions
  for Sums of Independent Random Variables}}}},\ \bibinfo {edition} {revised}\
  ed.\ (\bibinfo  {publisher} {Addison-Wesley},\ \bibinfo {address} {New
  York},\ \bibinfo {year} {1968})\BibitemShut {NoStop}%
\bibitem [{\citenamefont {Friedman}\ and\ \citenamefont
  {Martin}(1984)}]{Friedman1984p23}%
  \BibitemOpen
  \bibfield  {author} {\bibinfo {author} {\bibfnamefont {B.}~\bibnamefont
  {Friedman}}\ and\ \bibinfo {author} {\bibfnamefont {R.~F.}\ \bibnamefont
  {Martin}, \bibfnamefont {Jr.}},\ }\href@noop {} {\bibfield  {journal}
  {\bibinfo  {journal} {Physics Letters A}\ }\textbf {\bibinfo {volume}
  {105}},\ \bibinfo {pages} {23} (\bibinfo {year} {1984})}\BibitemShut
  {NoStop}%
\bibitem [{\citenamefont {Bouchaud}\ and\ \citenamefont
  {Le~Doussal}(1985)}]{Bouchaud:1985p7680}%
  \BibitemOpen
  \bibfield  {author} {\bibinfo {author} {\bibfnamefont {J.-P.}\ \bibnamefont
  {Bouchaud}}\ and\ \bibinfo {author} {\bibfnamefont {P.}~\bibnamefont
  {Le~Doussal}},\ }\href@noop {} {\bibfield  {journal} {\bibinfo  {journal}
  {Journal of Statistical Physics}\ }\textbf {\bibinfo {volume} {41}},\
  \bibinfo {pages} {225} (\bibinfo {year} {1985})}\BibitemShut {NoStop}%
\bibitem [{\citenamefont {Zacherl}\ \emph {et~al.}(1986)\citenamefont
  {Zacherl}, \citenamefont {Geisel}, \citenamefont {Nierwetberg},\ and\
  \citenamefont {Radons}}]{Zacherl:1986p7768}%
  \BibitemOpen
  \bibfield  {author} {\bibinfo {author} {\bibfnamefont {A.}~\bibnamefont
  {Zacherl}}, \bibinfo {author} {\bibfnamefont {T.}~\bibnamefont {Geisel}},
  \bibinfo {author} {\bibfnamefont {J.}~\bibnamefont {Nierwetberg}}, \ and\
  \bibinfo {author} {\bibfnamefont {G.}~\bibnamefont {Radons}},\ }\href@noop {}
  {\bibfield  {journal} {\bibinfo  {journal} {Physics Letters A}\ }\textbf
  {\bibinfo {volume} {114}},\ \bibinfo {pages} {317} (\bibinfo {year}
  {1986})}\BibitemShut {NoStop}%
\bibitem [{\citenamefont {Geisel}\ \emph {et~al.}(1985)\citenamefont {Geisel},
  \citenamefont {Nierwetberg},\ and\ \citenamefont
  {Zacherl}}]{Geisel:1985p8023}%
  \BibitemOpen
  \bibfield  {author} {\bibinfo {author} {\bibfnamefont {T.}~\bibnamefont
  {Geisel}}, \bibinfo {author} {\bibfnamefont {J.}~\bibnamefont {Nierwetberg}},
  \ and\ \bibinfo {author} {\bibfnamefont {A.}~\bibnamefont {Zacherl}},\
  }\href@noop {} {\bibfield  {journal} {\bibinfo  {journal} {Physical Review
  Letters}\ }\textbf {\bibinfo {volume} {54}},\ \bibinfo {pages} {616}
  (\bibinfo {year} {1985})}\BibitemShut {NoStop}%
\bibitem [{\citenamefont {Bleher}(1992)}]{Bleher:1992JSP315}%
  \BibitemOpen
  \bibfield  {author} {\bibinfo {author} {\bibfnamefont {P.~M.}\ \bibnamefont
  {Bleher}},\ }\href@noop {} {\bibfield  {journal} {\bibinfo  {journal}
  {Journal of Statistical Physics}\ }\textbf {\bibinfo {volume} {66}},\
  \bibinfo {pages} {315} (\bibinfo {year} {1992})}\BibitemShut {NoStop}%
\bibitem [{\citenamefont {Sz{\'a}sz}\ and\ \citenamefont
  {Varj{\'u}}(2007)}]{Szasz:2007v129p59}%
  \BibitemOpen
  \bibfield  {author} {\bibinfo {author} {\bibfnamefont {D.}~\bibnamefont
  {Sz{\'a}sz}}\ and\ \bibinfo {author} {\bibfnamefont {T.}~\bibnamefont
  {Varj{\'u}}},\ }\href@noop {} {\bibfield  {journal} {\bibinfo  {journal}
  {Journal of Statistical Physics}\ }\textbf {\bibinfo {volume} {129}},\
  \bibinfo {pages} {59} (\bibinfo {year} {2007})}\BibitemShut {NoStop}%
\bibitem [{\citenamefont {Dolgopyat}\ and\ \citenamefont
  {Chernov}(2009)}]{Dolgopyat:2009p16456}%
  \BibitemOpen
  \bibfield  {author} {\bibinfo {author} {\bibfnamefont {D.~I.}\ \bibnamefont
  {Dolgopyat}}\ and\ \bibinfo {author} {\bibfnamefont {N.~I.}\ \bibnamefont
  {Chernov}},\ }\href@noop {} {\bibfield  {journal} {\bibinfo  {journal}
  {Russian Mathematical Surveys}\ }\textbf {\bibinfo {volume} {64}},\ \bibinfo
  {pages} {651} (\bibinfo {year} {2009})}\BibitemShut {NoStop}%
\bibitem [{\citenamefont {Machta}\ and\ \citenamefont
  {Zwanzig}(1983)}]{Machta:1983p182}%
  \BibitemOpen
  \bibfield  {author} {\bibinfo {author} {\bibfnamefont {J.}~\bibnamefont
  {Machta}}\ and\ \bibinfo {author} {\bibfnamefont {R.}~\bibnamefont
  {Zwanzig}},\ }\href@noop {} {\bibfield  {journal} {\bibinfo  {journal}
  {Physical Review Letters}\ }\textbf {\bibinfo {volume} {50}},\ \bibinfo
  {pages} {1959} (\bibinfo {year} {1983})}\BibitemShut {NoStop}%
\bibitem [{\citenamefont {Klafter}\ \emph {et~al.}(1987)\citenamefont
  {Klafter}, \citenamefont {Blumen},\ and\ \citenamefont
  {Shlesinger}}]{Klafter:1987Stochastic}%
  \BibitemOpen
  \bibfield  {author} {\bibinfo {author} {\bibfnamefont {J.}~\bibnamefont
  {Klafter}}, \bibinfo {author} {\bibfnamefont {A.}~\bibnamefont {Blumen}}, \
  and\ \bibinfo {author} {\bibfnamefont {M.~F.}\ \bibnamefont {Shlesinger}},\
  }\href@noop {} {\bibfield  {journal} {\bibinfo  {journal} {Physical Review
  A}\ }\textbf {\bibinfo {volume} {35}},\ \bibinfo {pages} {3081} (\bibinfo
  {year} {1987})}\BibitemShut {NoStop}%
\bibitem [{\citenamefont {Zumofen}\ and\ \citenamefont
  {Klafter}(1993)}]{Zumofen:1993p804}%
  \BibitemOpen
  \bibfield  {author} {\bibinfo {author} {\bibfnamefont {G.}~\bibnamefont
  {Zumofen}}\ and\ \bibinfo {author} {\bibfnamefont {J.}~\bibnamefont
  {Klafter}},\ }\href@noop {} {\bibfield  {journal} {\bibinfo  {journal}
  {Physical Review E}\ }\textbf {\bibinfo {volume} {47}},\ \bibinfo {pages}
  {851} (\bibinfo {year} {1993})}\BibitemShut {NoStop}%
\bibitem [{\citenamefont {Kenkre}\ \emph {et~al.}(1973)\citenamefont {Kenkre},
  \citenamefont {Montroll},\ and\ \citenamefont {Shlesinger}}]{Kenkre:1973em}%
  \BibitemOpen
  \bibfield  {author} {\bibinfo {author} {\bibfnamefont {V.~M.}\ \bibnamefont
  {Kenkre}}, \bibinfo {author} {\bibfnamefont {E.~W.}\ \bibnamefont
  {Montroll}}, \ and\ \bibinfo {author} {\bibfnamefont {M.~F.}\ \bibnamefont
  {Shlesinger}},\ }\href@noop {} {\bibfield  {journal} {\bibinfo  {journal}
  {Journal of Statistical Physics}\ }\textbf {\bibinfo {volume} {9}},\ \bibinfo
  {pages} {45} (\bibinfo {year} {1973})}\BibitemShut {NoStop}%
\bibitem [{\citenamefont {Landman}\ \emph {et~al.}(1977)\citenamefont
  {Landman}, \citenamefont {Montroll},\ and\ \citenamefont
  {Shlesinger}}]{Landman:1977PNAS430}%
  \BibitemOpen
  \bibfield  {author} {\bibinfo {author} {\bibfnamefont {U.}~\bibnamefont
  {Landman}}, \bibinfo {author} {\bibfnamefont {E.~W.}\ \bibnamefont
  {Montroll}}, \ and\ \bibinfo {author} {\bibfnamefont {M.~F.}\ \bibnamefont
  {Shlesinger}},\ }\href@noop {} {\bibfield  {journal} {\bibinfo  {journal}
  {Proceedings of the National Academy of Sciences of the United States of
  America}\ }\textbf {\bibinfo {volume} {74}},\ \bibinfo {pages} {430}
  (\bibinfo {year} {1977})}\BibitemShut {NoStop}%
\bibitem [{\citenamefont {Cristadoro}\ \emph
  {et~al.}(2014{\natexlab{a}})\citenamefont {Cristadoro}, \citenamefont
  {Gilbert}, \citenamefont {Lenci},\ and\ \citenamefont
  {Sanders}}]{cristadoro:2014transport}%
  \BibitemOpen
  \bibfield  {author} {\bibinfo {author} {\bibfnamefont {G.}~\bibnamefont
  {Cristadoro}}, \bibinfo {author} {\bibfnamefont {T.}~\bibnamefont {Gilbert}},
  \bibinfo {author} {\bibfnamefont {M.}~\bibnamefont {Lenci}}, \ and\ \bibinfo
  {author} {\bibfnamefont {D.~P.}\ \bibnamefont {Sanders}},\ }\href@noop {}
  {\bibfield  {journal} {\bibinfo  {journal} {arXiv:1407.0227}\ } (\bibinfo
  {year} {2014}{\natexlab{a}})}\BibitemShut {NoStop}%
\bibitem [{Note1()}]{Note1}%
  \BibitemOpen
  \bibinfo {note} {Without loss of generality, we assume $\rho \geq \rho _\odot
  $. For definiteness of a unique relevant timescale, Eq.~\protect \textup
  {\hbox {\mathsurround \z@ \protect \normalfont (\ignorespaces \ref
  {eq:taur}\unskip \@@italiccorr )}}, the separation between the central disc
  and the outer ones should also be substantially larger than the separation
  between two outer discs.}\BibitemShut {Stop}%
\bibitem [{\citenamefont {Bunimovich}\ \emph {et~al.}(1991)\citenamefont
  {Bunimovich}, \citenamefont {Sinai},\ and\ \citenamefont
  {Chernov}}]{Bunimovich:1991p47}%
  \BibitemOpen
  \bibfield  {author} {\bibinfo {author} {\bibfnamefont {L.~A.}\ \bibnamefont
  {Bunimovich}}, \bibinfo {author} {\bibfnamefont {Y.~G.}\ \bibnamefont
  {Sinai}}, \ and\ \bibinfo {author} {\bibfnamefont {N.}~\bibnamefont
  {Chernov}},\ }\href@noop {} {\bibfield  {journal} {\bibinfo  {journal}
  {Russian Mathematical Surveys}\ }\textbf {\bibinfo {volume} {46}},\ \bibinfo
  {pages} {47} (\bibinfo {year} {1991})}\BibitemShut {NoStop}%
\bibitem [{\citenamefont {Gaspard}\ and\ \citenamefont
  {Gilbert}(2012)}]{Gaspard:2012p26117}%
  \BibitemOpen
  \bibfield  {author} {\bibinfo {author} {\bibfnamefont {P.}~\bibnamefont
  {Gaspard}}\ and\ \bibinfo {author} {\bibfnamefont {T.}~\bibnamefont
  {Gilbert}},\ }\href@noop {} {\bibfield  {journal} {\bibinfo  {journal}
  {Chaos}\ }\textbf {\bibinfo {volume} {22}},\ \bibinfo {pages} {026117}
  (\bibinfo {year} {2012})}\BibitemShut {NoStop}%
\bibitem [{\citenamefont {Chernov}(1997)}]{Chernov:1997p1}%
  \BibitemOpen
  \bibfield  {author} {\bibinfo {author} {\bibfnamefont {N.}~\bibnamefont
  {Chernov}},\ }\href@noop {} {\bibfield  {journal} {\bibinfo  {journal}
  {Journal of Statistical Physics}\ }\textbf {\bibinfo {volume} {88}},\
  \bibinfo {pages} {1} (\bibinfo {year} {1997})}\BibitemShut {NoStop}%
\bibitem [{\citenamefont {Gilbert}\ and\ \citenamefont
  {Sanders}(2009)}]{Gilbert:2009p3207}%
  \BibitemOpen
  \bibfield  {author} {\bibinfo {author} {\bibfnamefont {T.}~\bibnamefont
  {Gilbert}}\ and\ \bibinfo {author} {\bibfnamefont {D.~P.}\ \bibnamefont
  {Sanders}},\ }\href@noop {} {\bibfield  {journal} {\bibinfo  {journal}
  {Physical Review E}\ }\textbf {\bibinfo {volume} {80}},\ \bibinfo {pages}
  {041121} (\bibinfo {year} {2009})}\BibitemShut {NoStop}%
\bibitem [{Note2()}]{Note2}%
  \BibitemOpen
  \bibinfo {note} {Strictly speaking, the residence time, Eq.~\protect \textup
  {\hbox {\mathsurround \z@ \protect \normalfont (\ignorespaces \ref
  {eq:taur}\unskip \@@italiccorr )}}, accounts for the possibility of
  collisionless motion inside a cell. However in the narrow-corridor regime,
  the difference between $\tau _\protect \mathrm {R}$ and the actual residence
  time conditioned on particles performing collisions inside the cell is next
  order in the small parameter $\delta /\ell $ and will thus be
  neglected.}\BibitemShut {Stop}%
\bibitem [{Note3()}]{Note3}%
  \BibitemOpen
  \bibinfo {note} {The second moment actually scales with coefficient given by
  twice the Gaussian value. See discussion in Ref.~\cite
  {cristadoro:2014measuring}.}\BibitemShut {Stop}%
\bibitem [{\citenamefont {Cristadoro}\ \emph
  {et~al.}(2014{\natexlab{b}})\citenamefont {Cristadoro}, \citenamefont
  {Gilbert}, \citenamefont {Lenci},\ and\ \citenamefont
  {Sanders}}]{cristadoro:2014measuring}%
  \BibitemOpen
  \bibfield  {author} {\bibinfo {author} {\bibfnamefont {G.}~\bibnamefont
  {Cristadoro}}, \bibinfo {author} {\bibfnamefont {T.}~\bibnamefont {Gilbert}},
  \bibinfo {author} {\bibfnamefont {M.}~\bibnamefont {Lenci}}, \ and\ \bibinfo
  {author} {\bibfnamefont {D.~P.}\ \bibnamefont {Sanders}},\ }\href@noop {}
  {\bibfield  {journal} {\bibinfo  {journal} {Physical Review E, in press;
  arXiv:1405.0975}\ } (\bibinfo {year} {2014}{\natexlab{b}})}\BibitemShut
  {NoStop}%
\end{thebibliography}

%

\end{document}